\documentclass[aps,prd, onecolumn,showpacs,showkeys,amsmath,amssymb,
groupedaddress]{revtex4}
\usepackage{ifthen}
\newboolean{prd}
\setboolean{prd}{false}
\newboolean{arxiv}
\setboolean{arxiv}{true}
\newboolean{notprd}
\setboolean{notprd}{true}
\ifprd
\setboolean{notprd}{false}
\fi
\newboolean{notarxiv}
\setboolean{notarxiv}{true}
\ifarxiv
\setboolean{notarxiv}{false}
\fi
\usepackage{graphicx}
\usepackage[usenames]{xcolor}
\usepackage{subfigure}
\usepackage{amssymb}
\usepackage[latin1]{inputenc}
\usepackage{amsfonts}
\usepackage{setspace}
\usepackage{amsmath}
\usepackage{rotating}
\usepackage{ulem}
\usepackage{color}
\xdefinecolor{mylinkcolor}{rgb}{0,0,0.5}
\usepackage[
	bookmarksnumbered, bookmarksopen, bookmarksopenlevel=2,
	breaklinks=true,
	colorlinks=true, filecolor=mylinkcolor, citecolor=mylinkcolor,
	linkcolor=mylinkcolor, urlcolor=mylinkcolor, menucolor=mylinkcolor,
]{hyperref}

\usepackage{wasysym}
\newcommand{\be}{\begin{equation}}
\newcommand{\ee}{\end{equation}}
\newcommand{\bea}{\begin{eqnarray}}
\newcommand{\eea}{\end{eqnarray}}

\def\eq#1{{Eq.~(\ref{#1})}}

\allowdisplaybreaks
\begin{document}

\hypersetup{
	pdftitle={},
	pdfauthor={}
}

%\title{Linear stability analysis of general relativistic black hole 
%accretion and emergent acoustic geometry: I - spherically 
%symmetric accretion in a static spherically symmetric metric}

\title{Entropy of a generic null surface from its associated Virasoro algebra }
\author{Sumanta Chakraborty}
\email{sumanta@iucaa.in, sumantac.physics@gmail.com}
\author{Sourav Bhattacharya}
\email{sbhatta@iucaa.in}
\author{T.~Padmanabhan}
\email{paddy@iucaa.in}
\affiliation{IUCAA, Post Bag 4, Ganeshkhind, Pune University Campus, Pune 411 007, India}
% \\ Emails:~,~, 
\date{\today}
%%%%%%%%%%%%%%%%%%%%%%%%%%%%%%%%%%%%%%%%%%%%%%%%%%%%%%%%%%%%%%%%%%%%%%%%%%%%%%%%%%%%%%%%%%%%%%%%%%%
%%%%%%%%%%%%%%%%%%%%%%%%%%%%%%%%%%%%%%%%%%%%%%%%%%%%%%%%%%%%%%%%%%%%%%%%%%%%%%%%%%%%%%%%%%%%%%%%%%%
%%%%%%%%%%%%%%%%%%%%%%%%%%%%%%%%%%%%%%%%%%%%%%%%%%%%%%%%%%%%%%%%%%%%%%%%%%%%%%%%%%%%%%%%%%%%%%%%%%%
\begin{abstract}
\noindent
Null surfaces act as one-way membranes, blocking information from those observers who do not cross them (e.g., in the black hole and the Rindler spacetimes) and these observers associate an entropy (and temperature) with the null surface. The black hole entropy can be computed from the central charge of an appropriately defined, local, Virasoro algebra on the horizon. We show that one can extend these ideas to a general class of null surfaces, all of which possess a Virasoro algebra and a central charge, leading to an entropy density (i.e., per unit area) which is just $(1/4)$.  All the previously known results of associating entropy with horizons arise as special cases of this  very general property of null surfaces demonstrated here and we believe this work represents the derivation of the entropy-area law in the most general context. The implications are discussed.
\end{abstract}
%%%%%%%%%%%%%%%%%%%%%%%%%%%%%%%%%%%%%%%%%%%%%%%%%%%%%%%%%%%%%%%%%%%%%%%%%%%%%%%%%%%%%%%%%%%%%%%%%%%
%%%%%%%%%%%%%%%%%%%%%%%%%%%%%%%%%%%%%%%%%%%%%%%%%%%%%%%%%%%%%%%%%%%%%%%%%%%%%%%%%%%%%%%%%%%%%%%%%%%
%%%%%%%%%%%%%%%%%%%%%%%%%%%%%%%%%%%%%%%%%%%%%%%%%%%%%%%%%%%%%%%%%%%%%%%%%%%%%%%%%%%%%%%%%%%%%%%%%%%
\maketitle
%%%%%%%%%%%%%%%%%%%%%%%%%%%%%%%%%%%%%%%%%%%%%%%%%%%%%%%%%%%%%%%%%%%%%%%%%%%%%%%%%%%%%%%%%%%%%%%%%%%
%%%%%%%%%%%%%%%%%%%%%%%%%%%%%%%%%%%%%%%%%%%%%%%%%%%%%%%%%%%%%%%%%%%%%%%%%%%%%%%%%%%%%%%%%%%%%%%%%%%
%%%%%%%%%%%%%%%%%%%%%%%%%%%%%%%%%%%%%%%%%%%%%%%%%%%%%%%%%%%%%%%%%%%%%%%%%%%%%%%%%%%%%%%%%%%%%%%%%%%
\section{Introduction} 
\noindent
It is well known that one can associate  thermodynamic variables like entropy $(S)$ and temperature $(T)$ with null surfaces, that are perceived to be one-way membranes, by the class of observers who do not cross them (see e.g., Chapter 8 of \cite{Padmanabhan}). For example, observers at constant spatial coordinates located at $r>2M$ in the Schwarzschild metric will associate a temperature $T=(1/8\pi M)$  with the black hole horizon \cite{Bekenstein1,Bekenstein2,Bekenstein3,Hawking2} at $r=2M$. Similarly, an observer at constant spatial coordinates at $x>0$ in a Rindler spacetime (with the metric $ds^2 = -g^2 x^2 dt^2 + d{\mathbf x}^2$)
will associate a temperature $T=(g/2\pi)$ with the Rindler horizon \cite{Davies,Unruh:1976db}, at $x=0$. Both these observers will associate an entropy density (entropy per unit area) of $(1/4)$ with the respective horizons. (In fact, the near horizon metric of the Schwarzschild black hole can be reduced to the form of the Rindler metric; this itself suggests that, as long as physical phenomena are reasonably local, we should get similar results in these two cases.) The freely falling observers in either spacetime will not perceive the relevant null surface ($r=2M$ in the black hole case and $x=0$ in the Rindler case) as endowed with thermodynamic properties, because these observers will eventually cross these null surfaces; therefore, these null surfaces do not act as one-way membranes for these observers, see \cite{Padmanabhan1,Padmanabhan:2013nxa,Jacobson:2003wv} and also references therein. 

For a broad class of, rather generic, null surfaces in an arbitrary spacetime, one can introduce a set of observers who do not cross these null surfaces and perceive them as one-way membranes. It seems reasonable to assume that physics should be local and the behaviour in a local region around an event should depend only on the geometrical properties around that event. It is also widely accepted that black hole horizon has an entropy density of 1/4 per unit area. An observer close the the horizon will see that the metric is well approximated by Rindler metric and locality requires a patch of this Rindler metric to have the entropy density 1/4. It is, therefore, important to investigate whether, in such a general context,  these observers will associate thermodynamic variables with such a generic 
null surface. Earlier work has shown  that there exist several deep connections between the properties of null surfaces and gravitational dynamics which suggest that this could be true \cite{Padmanabhan1,Padmanabhan:2013nxa}. If so, then we can obtain a  unified picture of the connection between thermodynamics and null surfaces, \textit{with all the previously known cases being reduced to just special cases of this result}. 
  
In this work, we will show that this is indeed the case. We will demonstrate that one can associate a very natural Virasoro algebra (e.g. \cite{Zwiebach}) with a general class of null surfaces. This Virasoro algebra has a central charge which, through Cardy's formula \cite{Cardy1,Cardy2}, leads to the entropy of the null surface. The explicit calculation shows that the entropy per unit area is $(1/4)$ which is consistent with the standard results for black  hole horizons, cosmological horizons etc. 

This work is based on the approach pioneered by Carlip \cite{Carlip,Carlip:2014pma} who first proved similar results in the specific case of a black hole horizon (see also \cite{Strominger:1997eq}) and the covariant phase space method (see \cite{Hajian:2015xlp} for a recent review and references therein). To obtain the  general result of this paper, we  add certain new ingredients and modify this approach suitably. We shall set $c=1=G$ throughout and will work with the mostly positive signature of the metric.
%%%%%%%%%%%%%%%%%%%%%%%%%%%%%%%%%%%%%%%%%%%%%%%%%%%%%%%%%%%%%%%%%%%%%%%%%%%%%%%%%%%%%%%%%%%%%%%%%%%
%%%%%%%%%%%%%%%%%%%%%%%%%%%%%%%%%%%%%%%%%%%%%%%%%%%%%%%%%%%%%%%%%%%%%%%%%%%%%%%%%%%%%%%%%%%%%%%%%%%
%%%%%%%%%%%%%%%%%%%%%%%%%%%%%%%%%%%%%%%%%%%%%%%%%%%%%%%%%%%%%%%%%%%%%%%%%%%%%%%%%%%%%%%%%%%%%%%%%%%
\section{The formalism} 
 \noindent
It is known that the Bekenstein entropy can be derived just by exploiting the near horizon conformal properties \cite{Carlip,Carlip:2014pma,Strominger:1997eq,Dreyer:2001py,Silva:2002jq,Carlip:2007qh,Dreyer:2013noa,Kang:2004js}. This procedure imposes some suitable fall-off conditions on the metric near the horizon  and investigates the symmetries that preserve those fall-off conditions. The algebra of the conserved charges corresponding to these symmetries turns out to be a Virasoro algebra with a central extension. Then, using Cardy's formula \cite{Cardy1,Cardy2}, we can compute the entropy of the spacetime from the central extension. The result agrees with the standard expression for entropy, viz., one-quarter per unit area. (For a review of this approach and references, see, e.g., \cite{Carlip:2014pma}.) Thus the entropy of a black hole spacetime can be determined purely from the local symmetries at its boundary, i.e., at the event horizon. We shall make these ideas  more explicit in the sequel. 

A new ingredient was added to this approach couple of years back which further emphasized the local nature of this procedure. In this approach \cite{Majhi:2011ws,Majhi:2012tf,Majhi:2013jpk,Majhi:2014lka}, one uses diffeomorphism invariance of  the  Gibbons-Hawking-York surface term in the gravitational action, along with the near horizon's symmetries, to obtain the Virasoro algebra and the entropy. We will be using this formalism in what follows. 

It is well known that~\cite{Padmanabhan}: (i) the Hilbert action does not possess a valid functional derivative with respect to the metric and (ii) this situation can be remedied by adding a suitable counter-term to the action. This counter-term contains the integral of the extrinsic curvature $K=-\nabla_a N^a$ over the boundary, which has an induced metric $h_{ab}$ and normal to the surface being $N^{a}$. The boundary integral could be converted to a volume integral via the Gauss theorem (with any suitable extension of $N^a$ away from the boundary):
%
%\begin{widetext}
\begin{eqnarray}
\int_{\rm Boundary} \sqrt{-h}\, d^3 x \,  K= \int \sqrt{-g}\, d^4 x \, \nabla_a(K N^a(x^{\mu})). 
\label{e1'}
\end{eqnarray}
%\end{widetext}
%
The invariance of this boundary term under the diffeomorphisms  gives rise to a conserved Noether charge. In particular, the Noether charge corresponding to coordinate transformations, related to  the local symmetries near the horizon, allows one to define a natural Virasoro algebra on the event horizon. The central extension of this Virasoro algebra leads to the correct entropy through Cardy's formula. This formalism was later successfully extended to cosmological black holes \cite{Majhi:2014hpa} and  black hole spacetimes endowed with cosmological event horizons \cite{Bhattacharya:2015mja}. 

In this context we would like to point out that the counter-term as introduced above is by no means unique~\cite{Charap:1982kn}-\cite{Barnich:2007bf}. However, it is well known that many  counter-terms differ only by terms that do not involve normal derivatives on the boundary. In other words, all the possible counter-terms must contain the same normal derivative terms, since that is the criterion that a counter-term must satisfy~\cite{Parattu:Chak}. Hence the leading order contributions from the counter-terms on the null surface would be same as that obtained from $K\sqrt{|h|}$, leading to similar conclusions.

The previous works we cited above, deal essentially with stationary black hole spacetimes having a Killing vector field which becomes null on the horizon. Our aim is to extend this procedure to a generic null surface which, of course, will not be a Killing horizon.  We shall consider a general null surface and describe it in terms of some suitable null coordinates, called the Gaussian null coordinates (see e.g., \cite{Chakraborty:2015hna, Parattu:2015gga}), defined locally in the vicinity of the surface, and use the spacetime geometry near that surface. Such surfaces, as we shall see below, are {\it much more general} than the Killing horizons. We shall then use the  surface term formalism \cite{Majhi:2011ws,Majhi:2012tf,Majhi:2013jpk,Majhi:2014lka} mentioned above to find the conserved charges corresponding to the local symmetry of the null surface geometry. We will  show that, for a wide class of null surfaces, satisfying rather mild physical requirements, we can define a natural Virasoro algebra. The 
central extension of this algebra will lead to the entropy.

The rest of the letter is organized as follows. In the next section we outline the essential technical tools required for our analysis. We derive the Virasoro algebra and the expression for the entropy in Sec. \ref{Entropy}. Finally we end with a summary and outlook of our result. 
%%%%%%%%%%%%%%%%%%%%%%%%%%%%%%%%%%%%%%%%%%%%%%%%%%%%%%%%%%%%%%%%%%%%%%%%%%%%%%%%%%%%%%%%%%%%%%%%%%%
%%%%%%%%%%%%%%%%%%%%%%%%%%%%%%%%%%%%%%%%%%%%%%%%%%%%%%%%%%%%%%%%%%%%%%%%%%%%%%%%%%%%%%%%%%%%%%%%%%%
%%%%%%%%%%%%%%%%%%%%%%%%%%%%%%%%%%%%%%%%%%%%%%%%%%%%%%%%%%%%%%%%%%%%%%%%%%%%%%%%%%%%%%%%%%%%%%%%%%%
\section{The geometry} 

The most general  geometry in the local neighborhood of a null surface can be described using the Gaussian Null Coordinates (see e.g., Ref. \cite{Chakraborty:2015hna,Parattu:2015gga} and references therein), in which the metric takes the form:
%
%\begin{widetext}
\begin{eqnarray}
ds^2=-2x\alpha du^2 -2dudx -2x\beta_idudx^i +q_{ij}dx^idx^j,
\label{e2}
\end{eqnarray}
%\end{widetext}
%
where $x^i~(i=1,2)$ are the coordinates on a spatial 2-dimensional surface. This surface is transported  along the vectors along $u$ and $x$. Since $g_{xx}=0$, the vector along $x$ ($\equiv \partial_x$) is null. The vector $\partial_u$ becomes null on $x=0$. The six functions $\alpha$, $\beta_i$ and $q_{ij}$ are assumed to be non-vanishing and well behaved but are arbitrary otherwise. They contain the $10-4=6$ functional degrees of freedom of the metric, showing that no generality has been lost by this coordinate choice. It is easy to check that the normal to the null surface, $\nabla_a (2 \alpha x)$, is also a null vector on $x=0$, with its norm vanishing as ${\cal O}(x)$.

We will now impose the following conditions on the metric: \textit{On the null surface}, $x=0$, the functions $\alpha$ and $q_{ij}$ are assumed to be independent of $u$ and one of the coordinates, say, $x^1$ and it is periodic. We stress that we impose no conditions on the spacetime away from the null surface and no conditions on $\beta_i$ anywhere. (If one thinks of $x^1$ as analogous to an angular coordinate and $u$ as analogous to time coordinate, our condition has a superficial similarity to stationarity and axisymmetry on $x=0$. It can be shown, however, that the null surface with these conditions is {\it not} a Killing horizon and that our geometry is much more general; a proof is provided in the Appendix of this letter.)
%%%%%%%%%%%%%%%%%%%%%%%%%%%%%%%%%%%%%%%%%%%%%%%%%%%%%%%%%%%%%%%%%%%%%%%%%%%%%%%%%%%%%%%%%%%%%%%%%%%
%%%%%%%%%%%%%%%%%%%%%%%%%%%%%%%%%%%%%%%%%%%%%%%%%%%%%%%%%%%%%%%%%%%%%%%%%%%%%%%%%%%%%%%%%%%%%%%%%%%
%%%%%%%%%%%%%%%%%%%%%%%%%%%%%%%%%%%%%%%%%%%%%%%%%%%%%%%%%%%%%%%%%%%%%%%%%%%%%%%%%%%%%%%%%%%%%%%%%%%
\section{The entropy}\label{Entropy}

We shall briefly outline the derivation of the Virasoro algebra and the entropy, omitting algebraic details, which are similar to those in previous works and  can be found in, e.g. \cite{Majhi:2012tf}. If $\zeta^a$ is a vector field that generates diffeomorphism, the invariance of the boundary term of the action, gives a conserved ($\nabla_a J^a=0$) Noether current $J^{a}=\nabla_b J^{ab}$ with:
\begin{equation}\label{Eq01}
J^{ab}[\zeta]=\frac{K}{8\pi }(\zeta^a N^b-\zeta^b N^a)                                                       
\end{equation} 
Note that the Noether potential introduced above has some ambiguities. For example, one can add any second rank antisymmetric tensor to the above equation, which would not affect the conservation of the Noether current anyway, but would render the Noether charges ambiguous~\cite{Lee:1990nz,Wald:1993nt,Iyer:1994ys}. However in the context of Rindler or Killing horizons it is well known that the correct Noether potential corresponds to the one presented in Eq.~(\ref{Eq01}), as explicitly shown in~\cite{Majhi:2012tf} (see also~\cite{Lee:1990nz}). Consequently, in this work, we will take the Noether potential to be given by the same expression in order to have the appropriate Rindler/Killing limit of our later calculations. Certainly, this is not the unique way to do so, but definitely seems to be a very reasonable one. Given the Noether potential, corresponding Noether charge is given by:
%
%\begin{widetext}
\begin{eqnarray}
Q[\zeta]=\frac12\int d\Sigma_a J^a=\frac12 \int dq_{ab} J^{ab}
\label{charge1}
\end{eqnarray}
%\end{widetext}
%
where $\Sigma$ is some suitable hypersurface and the last integral is  over the 2-surface on the boundary, which, for our case, would be the spatial 2-surface with metric $q_{ab}$ in~\eq{e2}. The area element on that surface is given by  $d q_{ab}= \sqrt{q}\,(N_a M_b-N_b M_a) d^2x^i$, where $N^a$ and $M^a$ are respectively the unit spacelike and timelike normals to it.
We take them to be :
\begin{eqnarray}
N^a&=& \left(\frac{1}{\sqrt{2 x\alpha}},\sqrt{2x\alpha},\frac{x\beta^i}{\sqrt{2x\alpha}}\right) \nonumber\\
M^a&=&\left(\frac{1}{\sqrt{2x\alpha}},0,0,0\right)
\label{e7}
\end{eqnarray}
so that :
%
%\begin{widetext}
\begin{eqnarray}
K=-\nabla_a N^a= -\sqrt{\alpha/2x} +{\cal O} (x)
\label{e8}
\end{eqnarray}
%\end{widetext}
%
The Lie bracket algebra of the Noether charges is given by :
%
%\begin{widetext}
\begin{eqnarray}
[Q[\zeta_m], Q[\zeta_n]]:= \frac12  \int d q_{ab} (\zeta_m^a J^b[\zeta_n]-\zeta_n^a J^b[\zeta_m])
\label{charge2}
\end{eqnarray}
We now have to choose appropriate diffeomorphism generators $\{\zeta_m\}$, such that we preserve the near null surface geometry of \eq{e2}. Since the essential feature of the null surface we want to capture, viz., the blocking of the information is solely determined by $g_{uu}$, $g_{ux}$ and $g_{xx}$, we need to pick up only those vector fields which acts as an isometries on these metric components at $x=0$. Setting $\pounds_{\zeta} g_{ab}= \zeta^c\partial_c g_{ab}+g_{c(a}\partial_{b)}\zeta^c=0$ for these components, we find that
%
%\begin{widetext}
\begin{eqnarray}
\zeta^a \equiv \{F(u,x^i), -x F'(u,x^i), 0,0 \}
\label{e9}
\end{eqnarray}
%\end{widetext}
%   
where a prime denotes differentiation with respect to $u$. We have set all transverse components of $\zeta^a$ to zero, because all such components give vanishing contribution, when substituted into \eq{charge1} and \eq{charge2}.

The Noether charge as presented in Eq.~(\ref{charge1}) actually corresponds to the charge associated with the varied metric, characterized by $\zeta ^{a}$. One can argue the integrability of the charges~\cite{Compere:2015knw}, as the charge variation essentially depends linearly on the function $F(u,x^i)$, which being a scalar function, leads to commuting variations guaranteeing the  integrability of the charges.

Following the standard procedure, we next find a set of vector fields which obey an infinite dimensional discrete Lie algebra on a circle on or around $x=0$. In order to do this, we expand the fields as $\zeta=\Sigma A_m \zeta_m$,
where $m$ are integers and $\zeta_m$'s are regarded as the mode functions. The desired Lie bracket algebra is then 
%
%\begin{widetext}
\begin{eqnarray}
[\zeta_m,\zeta_n]=-i(m-n)\zeta_{m+n}
\label{e10}
\end{eqnarray}
%\end{widetext}
%   
An obvious choice is $\zeta_m \equiv F(u, x^i)={\ell}^{-1}e^{im (\ell u+p_ix^i)}$, where $\ell$ is an arbitrary constant. We substitute this expression into \eq{charge1} and \eq{charge2}. Our assumption that neither $\alpha$ nor $\sqrt{q}$ depend upon the $u,~x^1$ on the null surface and imposing periodicity of $2\pi$ on $x^1$, gives us the conserved charges
%
%\begin{widetext}
\begin{eqnarray}
Q_m =\frac{1}{8\pi  \ell}\int d^{2}x\sqrt{q} \alpha\delta _{m,0}
\label{v1}
\end{eqnarray}
and their algebra
%\begin{widetext}
\begin{eqnarray}
[Q_m, Q_n]= -\frac{i}{8\pi \ell}\int d^{2}x\sqrt{q}\alpha (m-n)\delta _{m+n,0}-\frac{i\ell m^3}{16\pi }\int  d^{2}x\frac{\sqrt{q}}{\alpha}\delta _{m+n,0}
\label{v2}
\end{eqnarray}
%\end{widetext}
%   
This is clearly a Virasoro algebra with a central extension, and looks very similar to the one we get for the Killing horizons \cite{Majhi:2011ws} considered earlier in the literature. However, here $\alpha$ is {\it not} a constant, so cannot be pulled out of the integration unlike in the case when the null surface is a Killing horizon. However we can work with the \textit{densities} and easily obtain an expression for entropy density. We once again emphasize that the above derivation of the Virasoro algebra depends crucially on the fact that the co-dimension two surface at $x=0$ is compact (i.e., the $2\pi$ periodicity of $x^1$ above), for otherwise the indices on the functions determining the diffeomorphism generating vector field $\zeta^{a}$ would not be integer and hence the algebra would not exist. (Note, however, that
even when this surface is not compact, one can do this by imposing periodic boundary conditions
at a given length scale, which will render it effectively compact. In the spirit of local analysis we
are performing there will always be some suitable background length scale which can be used
for this purpose.)   Finally, $Q_{0}$ and the central charge $C$ for \eq{charge1} and \eq{charge2} are given by 
%
%\begin{widetext}
\begin{eqnarray}
Q_0 =\frac{1}{8\pi  \ell}\int d^{2}x\sqrt{q} \alpha, \quad C= \frac{3\ell }{4\pi }\int  d^{2}x\frac{\sqrt{q}}{\alpha}
\label{v4}
\end{eqnarray}
We now wish to apply Cardy's formula~\cite{Cardy1, Cardy2} to obtain the entropy density of our null surface. To do this note that a small area element $\Delta A\equiv d^{2}x$ will contribute $\Delta Q_0\equiv {\cal Q}_0\,\Delta A $ (to $Q_0$) and an amount 
$\Delta C\equiv {\cal C} \, \Delta A $ to the central charge where ${\cal Q}_0=\sqrt{q} \alpha/(8\pi  \ell)$ and ${\cal C}=3\ell \sqrt{q} /(4\pi \alpha)$ are the integrands in \eq{v4}. Using Cardy's formula, we associate with this area element $\Delta A$ the entropy 
\begin{eqnarray}
\Delta S=2\pi\, \sqrt{ \frac{ {\cal C}\, \Delta A {\cal Q}_0\, \Delta A}{6}}= \frac{\sqrt{q}}{4}\, \Delta A
\label{tpv5}
\end{eqnarray}
The crucial square root in Cardy's formula allows us to interpret the resulting expression in terms of an entropy density :
\begin{eqnarray}
s\equiv \frac{\Delta S}{\Delta A}= \sqrt{ \frac{ {\cal C} {\cal Q}_0}{6}}= \frac{\sqrt{q}}{4}
\label{v5}
\end{eqnarray}
Since we are considering a very general situation, we do not have a result equivalent to the constancy of $\alpha$, (which is the analogue of surface gravity)  on the null surface. Nevertheless, it is interesting that if we apply the Cardy formula to the contribution from each area element, we can  work with local densities and obtain the expected result. We find that even though the temperature associated with the null surface is not a constant, we  still obtain an appropriate entropy density. (We will comment on this fact right at the end.)
%%%%%%%%%%%%%%%%%%%%%%%%%%%%%%%%%%%%%%%%%%%%%%%%%%%%%%%%%%%%%%%%%%%%%%%%%%%%%%%%%%%%%%%%%%%%%%%%%%%
%%%%%%%%%%%%%%%%%%%%%%%%%%%%%%%%%%%%%%%%%%%%%%%%%%%%%%%%%%%%%%%%%%%%%%%%%%%%%%%%%%%%%%%%%%%%%%%%%%%
%%%%%%%%%%%%%%%%%%%%%%%%%%%%%%%%%%%%%%%%%%%%%%%%%%%%%%%%%%%%%%%%%%%%%%%%%%%%%%%%%%%%%%%%%%%%%%%%%%%
\section{Summary and outlook }
 
We believe this result is  important and provides a key ``missing link'' in the study of null surfaces vis-a-vis gravitational thermodynamics. This is mainly due to the following facts :

\noindent 1. As pointed out earlier in references~\cite{Majhi:2011ws, Majhi:2012tf}, the degrees of freedom associated with the entropy of null surfaces (including event horizons) are observer dependent. (This should have been obvious from the fact that  freely falling and stationary observers will attribute different thermodynamic properties to the black hole horizon.) The diffeomorphisms, relevant to the observer who perceives a null surface as a horizon, are those  which preserve the near horizon geometry. These are a subset of all possible diffeomorphisms. Therefore, such an observer cannot eliminate \textit{all} the gauge degrees of freedom by using all possible diffeomorphism. This results in the transmutation of some gauge degrees of freedom to physical degrees of freedom, as far as this particular class of observers is concerned. This links up the observer dependence of horizon thermodynamics to the structural features of the diffeomorphism.

\noindent 2. Our derivation is remarkably local. It seem reasonable that all physics, including thermodynamics of horizons, should have a proper local description because, operationally, all the relevant measurements will be local. The locality in our derivation is based on three facts: (i) We consider diffeomorphisms near the horizon and use only the structural features of the metric near and on the horizon. (ii) We use the behaviour of the boundary term in the action under  diffeomorphism with the boundary being the relevant null surface. Again, no bulk construction is required. (iii) We show that a local version of the Cardy formula does give the correct answer. (iv)  It should be stressed that our result is completely independent of the asymptotic structure of the spacetime.

\noindent 3. We have shown that when a null surface is perceived to be a one-way membrane by a particular congruence of observers, they will associate an entropy with it. This directly links inaccessibility of information with entropy, which is gratifying. Moreover, the result holds for a very wide class of null surfaces and also can be generalized to arbitrary spacetime dimensions, in a straightforward manner. We only needed to make minimal conditions on the metric, that too only \textit{on} the null surface. \textit{The spacetime is completely arbitrary away from the null surface.}

\noindent 4. All the previous results known in the literature in the context of black holes, cosmology, non-inertial frames etc. become just special cases of this very general result.  Such a unified perspective will be useful in further investigations. 

Incidentally, the analysis will \textit{not} go through if, on the null surface, $\alpha$ depends on  $u$. In that case, we find that the Virasoro algebra is  not closed. The full expression demonstrating this is rather involved, but this is even true if we just retain terms linear in $\partial_u\alpha$, where we obtain the commutator :
%
%\begin{widetext}
\begin{eqnarray}
[Q_{m},Q_{n}]=-\frac{i}{8\pi  \ell}\int d^{2}x\sqrt{q}\left(\alpha-\frac{\partial _{u}\alpha}{\alpha}\right)(m-n)\delta_{m+n,0}
-\frac{im^3 \ell}{16\pi }\int  d^{2}x\frac{\sqrt{q}}{\alpha} \left(1- \frac{\partial_u\alpha}{2\alpha^2}  \right)\delta_{m+n,0}
\label{v3}
\end{eqnarray}
%\end{widetext}
% 
while the Noether charge is given by \eq{v1} with $\alpha$ replaced by $\alpha -(\partial _{u}\alpha/2\alpha)$. (Note the factor $1/2$ which causes the problem.) Hence part of the algebra containing the charges does not close unless we set $\partial _{u}\alpha=0$ on the null surface. It is not clear whether this is fundamental restriction of merely a limitation of this approach.

At this outset, let us point out two important issues that one should keep in mind. The first one corresponds to the usage and validity of Cardy's formula. In particular, having a Virasoro algebra does not {\it a priori} ensure the validity of Cardy's formula,  a property of the two-dimensional Conformal Field Theory. Further, it requires a set of reasonable physical conditions: (a) Unitarity, (b) Modular Invariance and (c) Discrete Spectrum and it does not seem to be valid for small central charges~\cite{Loran:2010bd}. This issue is more general than those addressed in this paper and we hope
to return to it in a future work.  Secondly, even though works well for the near extremal cases (i.e., $\alpha$ is nonvanishing but otherwise arbitrarily small), our method is not adapted for exactly extremal null surfaces ($\alpha=0$). For this, some other techniques must be devised, e.g.~\cite{Hajian:2015eha, Hajian:2013lna}.  We, however, believe that there is a lot to be learnt from the structure of generic, non-extremal, case and our
approach is well-suited for it. 

We  emphasize  that the present work provides a {\it paradigm shift } from the existing literature~\cite{Carlip} -- by generalizing the formalism to an arbitrary  null surface. The conventional Killing horizons are {\it only subsets} of this much more general construction and hence \textit{all} the previous results are  certain special cases of our current proposal. In fact, in many physical situations, a Killing horizon will be an idealized concept. The null surfaces considered here are significantly more general (e.g., we are allowing non-constant surface gravity) and hence is closer to the realistic situations. Thus our work not only provides {\it the most general framework to date} to derive the entropy-area relation, and also probably reveals some deeper connection between the entropy/information, area and null surfaces in a much broader context in gravitational physics.

Finally we want to comment on the remarkable robustness of Cardy's formula in this context. The relation between the Virasoro algebra and the entropy (through Cardy formula) was originally derived in the context of flat spacetime physics~\cite{Cardy1, Cardy2}.  It is not obvious that the result should generalize to curved spacetime and black hole horizons, which was a  bit of a surprise in Carlip's approach~\cite{Carlip}. (This is possibly related to local nature of the result and the fact that, in the local inertial frame, we can use the special relativistic physics.) In obtaining the entropy \textit{density} in \eq{tpv5} for an arbitrary null surface, we have extended the applicability of Cardy formula even further. As we have shown, if one works with the densities $\mathcal{C}$ and $\mathcal{Q}_0$ corresponding to  $C$ and $Q_0$, we can reproduce the entropy \textit{density} correctly. It is important to understand why this approach works  and whether it can be derived from a more local procedure. This 
might, in turn, throw more light on the applicability of Cardy's formula in the context of curved geometry.
These issues are under investigation.
%%%%%%%%%%%%%%%%%%%%%%%%%%%%%%%%%%%%%%%%%%%%%%%%%%%%%%%%%%%%%%%%%%%%%%%%%%%%%%%%%%%%%%%%%%%%%%%%%%%
%%%%%%%%%%%%%%%%%%%%%%%%%%%%%%%%%%%%%%%%%%%%%%%%%%%%%%%%%%%%%%%%%%%%%%%%%%%%%%%%%%%%%%%%%%%%%%%%%%%
%%%%%%%%%%%%%%%%%%%%%%%%%%%%%%%%%%%%%%%%%%%%%%%%%%%%%%%%%%%%%%%%%%%%%%%%%%%%%%%%%%%%%%%%%%%%%%%%%%%
\vskip 1cm

\noindent
\textbf{Acknowledgement:} Research of SC is funded by  SPM fellowship from CSIR, Government
 of India. Research of TP is partially supported by J.~C.~Bose Research Grant, DST, Government of India. The authors wish to thank K.~Lochan
and K.~Parattu for useful discussions.
%%%%%%%%%%%%%%%%%%%%%%%%%%%%%%%%%%%%%%%%%%%%%%%%%%%%%%%%%%%%%%%%%%%%%%%%%%%%%%%%%%%%%%%%%%%%%%%%%%%
%%%%%%%%%%%%%%%%%%%%%%%%%%%%%%%%%%%%%%%%%%%%%%%%%%%%%%%%%%%%%%%%%%%%%%%%%%%%%%%%%%%%%%%%%%%%%%%%%%%
%%%%%%%%%%%%%%%%%%%%%%%%%%%%%%%%%%%%%%%%%%%%%%%%%%%%%%%%%%%%%%%%%%%%%%%%%%%%%%%%%%%%%%%%%%%%%%%%%%%
%%%%%%%%%%%%%%%%%%%%%%%%%%%%%%% NEW   %%%%%%%%%%%%%%%%%%%%%%%
\begin{center}
 {\bf Appendix} 
\end{center}
\noindent
For the sake of completeness, we show that the null surface we are dealing with is \textit{not} a Killing horizon. 

Let us first see when this null surface can correspond to a  Killing horizon. If we set $\beta_i=0$, $\alpha={\rm const.}$, take $q_{ij}$ to be independent of $u$ and define a timelike coordinate $t$ as $u=t-1/2\alpha \ln x$, then this metric takes  the familiar Rindler form, appropriate for a Killing horizon. The vector field $\partial_t$ is then a Killing vector and becomes null on the horizon ($x=0$) and the constant $\alpha$ is identified as the surface gravity.

 We shall pick up a subclass of~\eq{e2}, by restricting these functions  for our purpose, while still relaxing the Killing horizon conditions,  as follows.  
We assume that {\it on the null surface} $x=0$, both $\alpha$ and $q_{ij}$ are independent of $u$ and one of the $x^i$'s, say $x^1$.  We leave the functions $\beta_i$ arbitrary. We do not assume any specific functional forms of the metric functions for our current purpose.

The Lie derivative of the metric functions with respect to the vector field $\partial_u$ simply reads,
%
%\begin{widetext}
\begin{eqnarray}
\pounds_{\partial_u} g_{ab}= \partial_u g_{ab}.
\label{e3}
\end{eqnarray}
%\end{widetext}
%
From~\eq{e2}, and our imposed restrictions, the right hand side vanishes at least as ${\cal O}(x)$, as $x\to 0$. Thus the coordinate vector field $\partial_u$ becomes Killing as $x\to 0$. However,
it is certainly not orthogonal to spacelike hypersurfaces, as is evident from the metric. Does there exist some other vector field that is hypersurface orthogonal and simultaneously becomes Killing and null on $x=0$? In order to see this, we define two new basis vector fields for the `$u-x$' part of the metric,  
%
%\begin{widetext}
\begin{eqnarray}
   \widetilde{u}^a&=& \partial_u^a+(x\beta_1/q_{11})\partial_{x^1}^a+(x\beta_2/q_{22})\partial_{x^2}^a\nonumber\\ \widetilde{x}^a&=&\partial_x^a-(1/2x \alpha) \widetilde{u}^a
\label{e4}
\end{eqnarray}
%\end{widetext}
%
where we have used a orthogonal coordinate basis for the spatial metric $q_{ij}dx^idx^j$, for calculational convenience only.  It is easy to see that we now have an orthogonal basis for the spacetime~\eq{e2} : $(\widetilde{u}^a,\, \widetilde{x}^a,\, \partial_{x_1}^a,\,\partial_{x_2}^a)$, where the first 
two are not necessarily coordinate vector fields. The norms of $\widetilde{u}^a$ and  $\widetilde{x}^a$  are respectively given by, $ -2\alpha x + {\cal O} (x^2)$ and $(2\alpha x)^{-1}$. In this basis,
the metric can be reexpressed as
%
%\begin{widetext}
\begin{eqnarray}
g_{ab}= - (2\alpha x)^{-1} \widetilde{u}_a\widetilde{u}_b +(2\alpha x) \widetilde{x}_a \widetilde{x}_b +q_{ab}
\label{e5}
\end{eqnarray}
%\end{widetext}
%
The above metric certainly looks formally the same as that of near a Killing horizon. The vector field $\widetilde{u}^a$ is manifestly hypersurface orthogonal. Moreover, since we have seen that $\partial_u$ becomes Killing as $x\to 0 $, from the expression of $\widetilde{u}^a$, the first of \eq{e4}, we may expect it to become Killing there, as well. However, this is not the case, as can be seen below.     

If any vector field $v^a$ is Killing, we must have $\pounds_{v} g_{ab}=\nabla_{a}v_{b}+\nabla_bv_a=0$, for all the components of the metric. For $\widetilde{u}^a$, we find almost all components vanish on $x=0$
as ${\cal O}(x)$ except 
%
%\begin{widetext}
\begin{eqnarray}
\pounds_{\widetilde{u}}g_{ab}\vert_{a=x, b=i}= \beta_i
\label{	}
\end{eqnarray}
%\end{widetext}
%
Since $\beta_i$'s are arbitrary, it is evident that $\widetilde{u}^a$ is not a Killing vector field and hence the null surface $x=0$ is not a Killing horizon. Accordingly~\cite{Wald:1984rg}, $\alpha$ will not be a constant on that null surface (in particular, it would depend upon the transverse spatial coordinate, $x^2$, on $x=0$). Since $\widetilde{u}^a$ is manifestly hypersurface orthogonal and null on $x=0$, and as can be checked that the normal vector field $\nabla_a(2\alpha x)$ is null on $x=0$, it is clear that this null surface  can act as a one way membrane. 
%%%%%%%%%%%%%%%%%%%%%%%%%%%%%%%%%%%%%%%%%%%%%%%%%%%%%%%%%%%%%%%%%%%%%%%%%%%%%%%%%%%%%%%%%%%%%%%%%%%
%%%%%%%%%%%%%%%%%%%%%%%%%%%%%%%%%%%%%%%%%%%%%%%%%%%%%%%%%%%%%%%%%%%%%%%%%%%%%%%%%%%%%%%%%%%%%%%%%%%
%%%%%%%%%%%%%%%%%%%%%%%%%%%%%%%%%%%%%%%%%%%%%%%%%%%%%%%%%%%%%%%%%%%%%%%%%%%%%%%%%%%%%%%%%%%%%%%%%%%

%%%%%%%%%%%%%%%%%%%%%%%%%%%%%%%%%%%%%%%%%%%%%%%%%%%%%%%%%%%%%%%%%%%%%%%%%%%%%%%%%%%%%%%%%%%%%%%%%%%
%%%%%%%%%%%%%%%%%%%%%%%%%%%%%%%%%%%%%%%%%%%%%%%%%%%%%%%%%%%%%%%%%%%%%%%%%%%%%%%%%%%%%%%%%%%%%%%%%%%
%%%%%%%%%%%%%%%%%%%%%%%%%%%%%%%%%%%%%%%%%%%%%%%%%%%%%%%%%%%%%%%%%%%%%%%%%%%%%%%%%%%%%%%%%%%%%%%%%%%
%%%%%%%%%%%%%%%%%%%%%%%%%%%%%%%%%%%%%%%%%%%%%%%%%%%%%%%%%%%%%%%%%%%%%%%%%%%
\end{document}